\documentclass[pra, two column, show pacs]{revtex4}%
\usepackage{amsfonts}
\usepackage{amsmath}
\usepackage{amssymb}
\usepackage{graphicx}%
\providecommand{\U}[1]{\protect\rule{.1in}{.1in}}
\begin{document}
\title{Spin-sensitive atom mirror via spin-orbit interaction}
\author{Lu Zhou$^{1,3}$\footnote{lzhou@phy.ecnu.edu.cn }, Ren-Fei Zheng$^{1}$ and
Weiping Zhang$^{2,3}$}
\affiliation{$^{1}$Department of Physics, School of Physics and Material Science, East
China Normal University, Shanghai 200062, China}
\affiliation{$^{2}$Department of Physics and Astronomy, Shanghai Jiaotong University,
Shanghai 200240, China}
\affiliation{$^{3}$Collaborative Innovation Center of Extreme Optics, Shanxi University,
Taiyuan, Shanxi 030006, China}

\begin{abstract}
Based on the spin-orbit coupling which have been recently implemented in
neutral cold atom gas, here we propose a scheme to realize spin-dependent
scattering of cold atoms. In particular we consider a matter wave packet of
cold atom gas impinging upon a step potential created by the optical light
field, inside which the atoms are subject to spin-orbit interaction. We show
that the proposed system can act as a spin polarizer or spin-selective atom
mirror for the incident atomic beam. The principle and the operating parameter
regime of the system are carefully discussed.

\end{abstract}

\maketitle

\section{introduction}

Recent years have witnessed rapid development of atom interferometry, which
promises wide applications ranging from precision measurement to fundamental
quantum mechanics \cite{atom interferometry}. As one of the key elements in
atom interferometry, the technique of atom mirror have been experimentally
implemented and greatly improved along with the development of atom optics
\cite{atom optics}. The atom mirror was first proposed by Cook and Hill
\cite{first am theory}, and the very first experiment was carried out in the
late 1980s \cite{first atomic mirror}, in which the mirror reflection of a
thermal atomic beam was implemented. From then on, various kinds of atom
mirror have been developed with different purposes. Nowadays typical atom
mirror utilize the technique of bragg scattering \cite{bragg scattering} or
evanescent field formed at a dielectric interface.

It is well-known that the polarization of a circularly-polarized light will be
reversed upon reflection at normal incidence. From the quantum-mechanical
perspective, this is because each circularly-polarized photon is attributed
with a spin whose direction is parallel to that of the light propagation
\cite{circularly-polarized light}. In other words, the spin and center-of-mass
motion are attached to each other for circularly-polarized light. The light
polarization is an important physical quantity and can be modified during the
light propagation, which have many important applications such as polarized glasses.

Pseudo-spin can be constructed from the atomic internal energy level
structure, which provides an extra internal degree-of-freedom to the atomic
dynamics in analogy to the photon polarization. Only very recently have the
atomic spin been successfully attached to its center-of-mass motion via the
mechanism of artificial spin-orbit (SO) coupling \cite{dalibardRMP2011,1D-SO},
which also provides a new possibility to develop a spin-sensitive atom mirror.
Such an atom mirror can reflect the atomic beam and in the meanwhile reverse
its spin-polarization on demand, which can find interesting applications such
as atom interferometry with spin-dependent phase shifts \cite{spin-dependent
atom interferometry}.

In this paper we pay special attention to the case of the reflection of atoms
impinging upon a potential created by an off-resonant optical light field. The
atoms are subject to SO-coupling inside the potential, which is the origin of
the spin polarization of the reflected atomic beam. Previously the role of the
SO interaction on the tunnelling dynamics had been studied in
\cite{sablikovPRB2007,banPRA2012,juzeliunasPRL2008,zhangPRA2012,khodasPRL2004}%
. Our previous work also studied the Goos-H\"{a}nchen shifts in
spin-orbit-coupled cold atoms upon total-internal-reflection
\cite{zhouPRA2015}. However the role of SO-coupling on the atomic
spin-polarization properties upon reflection has not been explored to our
knowledge. In the following we will carefully explore how the atomic
populations are distributed among different spin states upon the reflection.
By analyzing these results, we discuss how to design an efficient
spin-sensitive atom mirror. Since SO-coupling is generated from a non-Abelian
gauge potential, we hope that our results can be useful for the design of
non-Abelian atom optics elements and atom interferometers which exploit the
non-Abelian Aharanov-Bohm effect \cite{non-Abelian}.

The paper is organized as follows: In Sec II we present our model and an
analyzation of the atomic scattering properties is presented. Sec III is
devoted to the calculation of the spin-polarization rate as well as the
reflectivity. Based on that, we discussed the role of the system acting as the
spin-sensitive atom mirror. A brief discussion on the case of one-dimensional
SO-coupling is given in Sec IV. Finally we conclude in Sec V.

\section{model}

\begin{figure}[h]
\includegraphics[width=8cm]{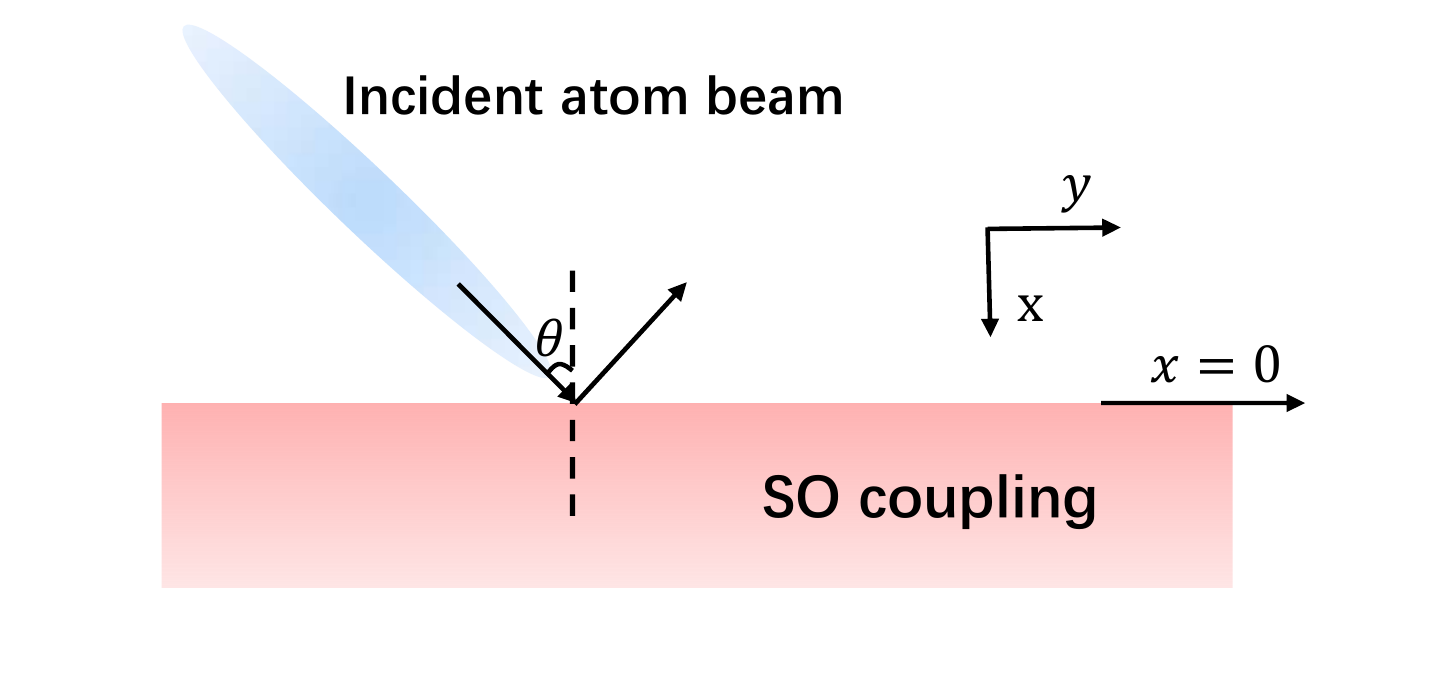}\caption{{\protect\footnotesize (Color
online) Schematic diagram showing the system under consideration. A beam of
cold atoms incident upon a potential barrier with an incident angle }$\theta
${\protect\footnotesize . As SO-coupling are superimposed inside the barrier,
the atoms will subject to spin-dependent scattering.}}%
\label{fig_scheme}%
\end{figure}

We consider the following model depicted in Fig. \ref{fig_scheme}: An atomic
beam incident upon a potential barrier from the $x$-direction. The scattering
potential is described by $V\left(  x\right)  =V_{0}\Theta(x)$, where
$\Theta(x)$ is the Heaviside step function. Such a step potential can be
created via a super-Gaussian laser beam with a large-enough order
\cite{SuperGaussian} and width compared to the atomic de-Broglie wavelength.
Inside the barrier, atoms are subject to Rashba SO-coupling, which can be
generated in neutral cold atoms with a tripod or ring coupling scheme
\cite{zhangPRL2012,JuzeliunasPRA2008,ZhangPRA2010,zhangRashba2015,ring-coupling}%
. The degree-of-freedom in $z$-direction is assumed to be frozen out by
external confinement, effectively reducing the dynamics to the $xy$-plane. The
effective atomic Hamiltonian inside the potential barrier can then be written
as
\begin{equation}
H=\frac{\hbar^{2}\boldsymbol{k}^{2}}{2m}+\frac{\hbar^{2}a}{m}\left(
k_{x}\sigma_{y}-k_{y}\sigma_{x}\right)  +V_{0}, \label{eq_hamiltonian}%
\end{equation}
where $\boldsymbol{k}^{2}=k_{x}^{2}+k_{y}^{2}$, and $a$, which characterizes
the strength of the SO coupling, is taken to be positive with dimension of wavenumber.

The eigenvectors of Hamiltonian (\ref{eq_hamiltonian}) split into two branches
and can generally be expressed as $\left\vert \phi_{\boldsymbol{k}}^{\pm
}\right\rangle =Ce^{i\left(  k_{x}x+k_{y}y\right)  }\left\vert \chi
_{\mathbf{k}}^{\pm}\right\rangle $ with $C$ the normalization constant and%
\begin{equation}
\left\vert \chi_{\mathbf{k}}^{\pm}\right\rangle =\binom{-2a\left(
k_{y}+ik_{x}\right)  }{E_{k}^{\pm}-V_{0}-\mathbf{k}^{2}}, \label{eq_phi}%
\end{equation}
in which $E_{\boldsymbol{k}}^{\pm}$ is the corresponding eigenenergy (scaled
by $\hbar^{2}/2m$) satisfying the relation
\begin{equation}
\left(  \boldsymbol{k}^{2}+V_{0}-E_{k}\right)  ^{2}-4a^{2}\boldsymbol{k}%
^{2}=0, \label{eq_energy}%
\end{equation}
from which we can get the energy spectrum $E_{k}=\mathbf{k}^{2}\pm2a\left\vert
\mathbf{k}\right\vert +V_{0}$.

In the situation considered here, the system is left free along the
$y$-direction and semi-infinite in the $x$-direction, thus $k_{y}$ is real and
$k_{x}$ is generally complex. As those had been illustrated in
Ref.~\cite{sablikovPRB2007}, the eigenfunctions of the system can be grouped
into three categories according to their properties: (i) propagating states
with $k_{x}$ real, (ii) evanescent states (only exist near the boundary of the
system and propagate along it) with $k_{x}=i\kappa$ (requiring $\left\vert
\kappa\right\vert <\left\vert k_{y}\right\vert $) and (iii) oscillating
evanescent states with $k_{x}=K_{x}^{\prime}+iK_{x}^{\prime\prime}$.

\begin{figure}[h]
\includegraphics[width=8cm]{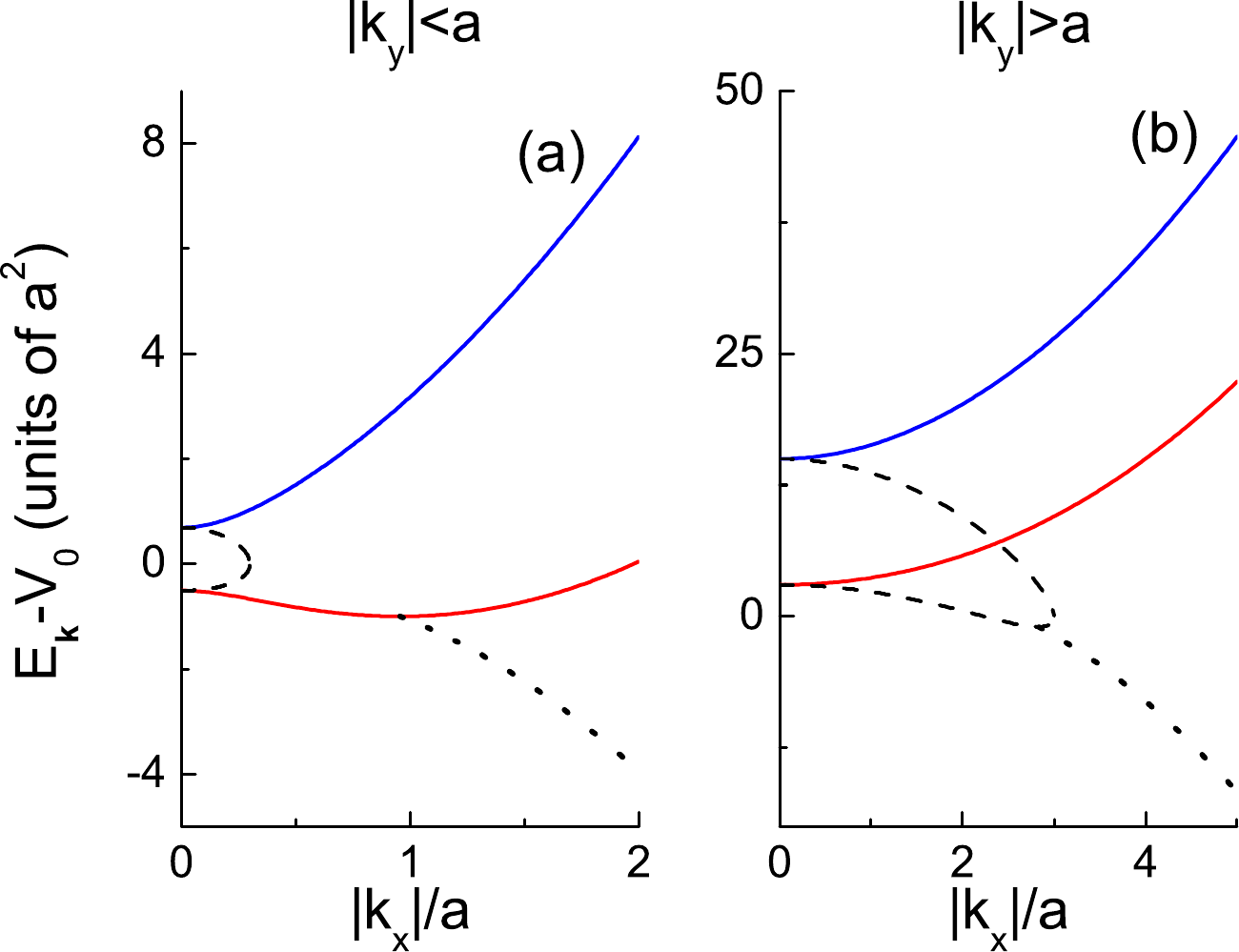}\caption{{\protect\footnotesize (Color
online) Typical energy spectra of the different states described by }%
$E_{k}=\mathbf{k}^{2}\pm2a\left\vert \mathbf{k}\right\vert +V_{0}%
${\protect\footnotesize . The solid (blue and red) lines correspond to the up
and down branches of the propagating states. The black dashed line represents
the evanescent states while the black dotted line is for the oscillating
evanescent states. (a) }$\left\vert k_{y}\right\vert <a$%
{\protect\footnotesize ; (b) }$\left\vert k_{y}\right\vert >a$%
{\protect\footnotesize .}}%
\label{fig_energy}%
\end{figure}

In order to better understand the properties of these eigenstates, we plot in
Fig. \ref{fig_energy} the energy spectra ($E_{\boldsymbol{k}} -V_{0}$) as a
function of $\left\vert k_{x}\right\vert $ for two typical values of $k_{y}$,
which exhibit different structures depending on the value of $k_{y}$. For both
cases, the two branches of propagating states are separated by a gap of $4 a
\left\vert k_{y}\right\vert $ at $\left\vert k_{x}\right\vert =0$. When
$\left\vert k_{y}\right\vert <a$, $\left\vert k_{x}\right\vert =\sqrt{a^{2}
-k_{y}^{2}}$ is the energy minimum of the lower propagating branch and the
dispersion curve of the evanescent states forms a lobe with its tip located at
$\left\vert k_{x}\right\vert =\left\vert k_{y}\right\vert $, which intersects
with the energy spectra of propagating states at $\left\vert k_{x}\right\vert
=0 $. While for $\left\vert k_{y}\right\vert >a$, $\left\vert k_{x}\right\vert
=0$ becomes the energy minimum of the lower propagating branch, which
intersects with the evanescent lobe at some finite $\left\vert k_{x}%
\right\vert $ besides $\left\vert k_{x}\right\vert =0$. The oscillating
evanescent states possess minimum energies among these three types of
solutions for both cases, and are linked to the energy minimum of the lower
propagating branch for $\left\vert k_{y}\right\vert <a $ and the evanescent
lobe for $\left\vert k_{y}\right\vert >a$.

A wide range of the atomic incident energy and incident angle will be
considered in the following discussion to include all three branches of the
atomic dispersion spectrum, under which circumstance the incident atom beam
will undergo either total internal reflection or partial reflection. We take
all these situations into account in order to find out the conditions under
which the spin-polarized reflection of atoms can be realized with high efficiency.

We assume that the atoms are initially prepared in the spin-$s$ ($s=\uparrow
,\downarrow$) propagating states with energy $E_{in}=k_{x}^{2}+k_{y}^{2}$ and
incident upon the step potential from $x<0$, then in the left region ($x<0$)
the wavefunction reads
\begin{equation}
\left\vert \psi_{\mathbf{k},s}^{\left(  L\right)  }\right\rangle =e^{i\left(
k_{x}x+k_{y}y\right)  }\left\vert s\right\rangle +e^{i\left(  -k_{x}%
x+k_{y}y\right)  }\sum\limits_{s^{\prime}}r_{ss^{\prime}}\left\vert s^{\prime
}\right\rangle , \label{eq_left}%
\end{equation}
in which $r_{ss^{\prime}}$ are reflection amplitudes.

From Fig. \ref{fig_energy} one can see that any equal-energy surface has two
intersections with the dispersion curve (including that of oscillating
evanescent eigenstates since $k_{x}=\pm K_{x}^{\prime}+iK_{x}^{\prime\prime}%
$), which indicate that the atom propagates inside the barrier in the form of
linear superpositions of two of the eigenfunctions described above. Depending
on the properties of the two eigenstates, the atom mirror operate in the
following four regimes:

(i) Two propagating states (2P). There exist two differenct cases in this
regime: One eigenstate is in the upper dispersion branch while the other one
is in the lower branch, or both eigenstates are in the lower dispersion
branch. For the first case the wavefunction in the right region ($x>0$) is%
\begin{equation}
\left\vert \psi_{\mathbf{k}}^{\left(  2P\right)  }\right\rangle =b_{1}%
e^{i\left(  k_{x1}x+k_{y}y\right)  }\left\vert \chi_{\mathbf{k}1}%
^{+}\right\rangle +b_{2}e^{i\left(  k_{x2}x+k_{y}y\right)  }\left\vert
\chi_{\mathbf{k}2}^{-}\right\rangle , \label{eq_2p_1}%
\end{equation}
with the modulus of the wave vector $\left\vert \mathbf{k}_{1}\right\vert
=-a+\sqrt{a^{2}+E_{in}-V_{0}}$ and $\left\vert \mathbf{k}_{2}\right\vert
=a+\sqrt{a^{2}+E_{in}-V_{0}}$. For the case of the two propagating eigenstates
both in the lower dispersion branch, which can only occur when $|k_{y}|<a$ as
shown in Fig. \ref{fig_energy}(a), one has negative $\partial E_{k}/\partial
k_{x}$ which means that the wave pointes outwards from the barrier while the
other one with positive $\partial E_{k}/\partial k_{x}$ propagate inwards
through the barrier. The corresponding wavefunction reads%
\begin{equation}
\left\vert \psi_{\mathbf{k}}^{\left(  2P\right)  }\right\rangle =b_{1}%
e^{i\left(  -k_{x1}x+k_{y}y\right)  }\left\vert \chi_{\mathbf{k}1}%
^{+}\right\rangle +b_{2}e^{i\left(  k_{x2}x+k_{y}y\right)  }\left\vert
\chi_{\mathbf{k}2}^{-}\right\rangle , \label{eq_2p_2}%
\end{equation}
with $\left\vert \mathbf{k}_{1}\right\vert =a-\sqrt{a^{2}+E_{in}-V_{0}}$ and
$\left\vert \mathbf{k}_{2}\right\vert $ kept unchanged.

(ii) One propagating state and one evanescent state (1P1E). When $\left\vert
\mathbf{k}_{1}\right\vert <\left\vert k_{y}\right\vert $ one wave becomes
evanescent and the wavefunction is%
\begin{equation}
\left\vert \psi_{\mathbf{k}}^{\left(  1P1E\right)  }\right\rangle
=b_{1}e^{-\kappa_{1}x+ik_{y}y}\left\vert \chi_{\mathbf{k}1}^{+}\right\rangle
+b_{2}e^{i\left(  k_{x2}x+k_{y}y\right)  }\left\vert \chi_{\mathbf{k}2}%
^{-}\right\rangle , \label{eq_1p1e}%
\end{equation}
with $\kappa_{1}=\sqrt{k_{y}^{2}-\left\vert \mathbf{k}_{1}\right\vert ^{2}}$.

(iii) Two evanescent states (2E). When $\left\vert \mathbf{k}_{2}\right\vert
<\left\vert k_{y}\right\vert $ both waves are evanescent with the wavefunction%
\begin{equation}
\left\vert \psi_{\mathbf{k}}^{(2E)}\right\rangle =b_{1}e^{-\kappa_{1}%
x+ik_{y}y}\left\vert \chi_{\mathbf{k}1}^{+}\right\rangle +b_{2}e^{-\kappa
_{2}x+ik_{y}y}\left\vert \chi_{\mathbf{k}2}^{-}\right\rangle , \label{eq_2e}%
\end{equation}
with $\kappa_{2}=\sqrt{k_{y}^{2}-\left\vert \mathbf{k}_{2}\right\vert ^{2}}$.

(iv) Two oscillating evanescent states (2OE). When $V_{0}>a^{2}+E_{in}$ inside
the barrier the atoms propagate with the wavevector $k_{x}=K_{x}^{\prime
}+iK_{x}^{\prime\prime}$ with $K_{x}^{\prime}$, $K_{x}^{\prime\prime}$ satisfy
$K_{x}^{\prime2}K_{x}^{\prime\prime2}=a^{2}\left(  V_{0}-E_{in}-a^{2}\right)
$ and $K_{x}^{\prime2}-K_{x}^{\prime\prime2}=2a^{2}+E_{in}-V_{0}-k_{y}^{2}$.
The wavefunction then reads%
\begin{align}
\left\vert \psi_{\mathbf{k}}^{(2E)}\right\rangle  &  =e^{-K_{x}^{\prime\prime
}x+ik_{y}y}\left[  b_{1}e^{iK_{x}^{\prime}x}\binom{a\left(  iK_{x}^{\prime
}-K_{x}^{\prime\prime}+k_{y}\right)  }{a^{2}+iK_{x}^{\prime}K_{x}%
^{\prime\prime}}\right. \nonumber\\
&  \left.  +b_{2}e^{-iK_{x}^{\prime}x}\binom{a\left(  -iK_{x}^{\prime}%
-K_{x}^{\prime\prime}+k_{y}\right)  }{a^{2}-iK_{x}^{\prime}K_{x}^{\prime
\prime}}\right]  , \label{eq_2oe}%
\end{align}

By integrating the Schr\"{o}dinger equation $H\psi(x)=E\psi(x)$ over the
interval expanded around the interface $x=0$, one can have%
\begin{equation}
-\frac{\partial\psi}{\partial x}+ia\sigma_{y}\psi|_{0^{+}}=-\frac{\partial
\psi}{\partial x}|_{0^{-}}. \label{eq_boundary}%
\end{equation}
The reflection amplitudes $r_{ss^{\prime}}$ can be determined from Eq.
(\ref{eq_boundary}) together with the continuous boundary condition. Due to
the symmetry inherent in the system, one can verify that $r_{\uparrow\uparrow
}\left(  k_{x},k_{y}\right)  =r_{\downarrow\downarrow}\left(  k_{x}%
,-k_{y}\right)  $ and $r_{\uparrow\downarrow}\left(  k_{x},k_{y}\right)
=-r_{\downarrow\uparrow}\left(  k_{x},-k_{y}\right)  $.

In the above analysis a sharp interface at $x=0$ is assumed to separate the
regions with and without the SO-interaction. This will require highly focused
dressing laser beam. In practical experiment a nonuniform SO-coupling strength
$a\left(  x\right)  $ with its interface extend over finite width $d$ would be
much easier to generate. A strict treatment on the case of general interface
will resort to time-dependent numerical simulation of the scattering dynamics,
which will be left for further investigation and not discussed here. However
we would like to note that, similar to the previous study on electron
scattering \cite{khodasPRL2004}, the essential physics derived here for the
case of sharp interface will be preserved for the case of a smooth adiabatic
interface $\lambda/d<<1$ ($\lambda=2\pi/\left\vert \mathbf{k}\right\vert $
characterize the atom wavelength). For a smooth interface $a\left(  x\right)
$ changes slowly on the scale of the atom wavelength and the atomic spin will
adjust itself adiabatically to the momentum when travelling through it. The
main physics of spin scattering will not be undermined with the simplified
treatment of a sharp interface.

\section{results and discussion}

Two physical quantities are at the core of the proposed spin-sensitive atom
mirror: Reflectivity and spin-polarization efficiency. For the present setup,
the spin-dependent reflectivity is defined as $r_{s}=\sum_{s^{\prime}%
}\left\vert r_{ss^{\prime}}\right\vert ^{2}$. The reflected particle current
in the longitudinal ($x$)-direction is $J_{s}\left(  k_{x},k_{y}\right)
=-\frac{\hbar k_{x}}{m}\sum\nolimits_{s^{\prime}}\left\vert r_{ss^{\prime}%
}\right\vert ^{2}$, while that for the spin current, we follow the definition
suggested by Shi \textit{et al}., \cite{ShiPRL2006}
\begin{align}
J_{s}^{j}\left(  k_{x},k_{y}\right)   &  =\left\langle v_{x}\sigma
_{j}\right\rangle \nonumber\\
&  =-\frac{\hbar k_{x}}{m}\left(
\begin{array}
[c]{c}%
2\operatorname{Re}\left(  r_{ss}r_{ss^{\prime}}^{\ast}\right) \\
-2\operatorname{Im}\left(  r_{ss}r_{ss^{\prime}}^{\ast}\right) \\
\left\vert r_{ss}\right\vert ^{2}-\left\vert r_{ss^{\prime}}\right\vert ^{2}%
\end{array}
\right)  . \label{eq_spin current}%
\end{align}
The efficiency of spin polarization upon reflection is characterized by the
ratio of the spin current to the particle current: $P_{s}^{j}=J_{s}^{j}/J_{s}%
$. Here we concentrate on the spin-$z$ component of the reflected current, and
the spin polarization rate upon reflection is characterized by the quantity
$P_{s}$, which is defined as%
\begin{equation}
P_{s}=\left(  1-J_{s}^{z}/J_{s}\right)  /2. \label{eq_polarization_efficiency}%
\end{equation}
If the spin of the atom completely flips upon reflection, $P_{s}=1$; Otherwise
if the atomic spin does not flip at all, $P_{s}=0$.

\begin{figure}[h]
\includegraphics[width=8cm]{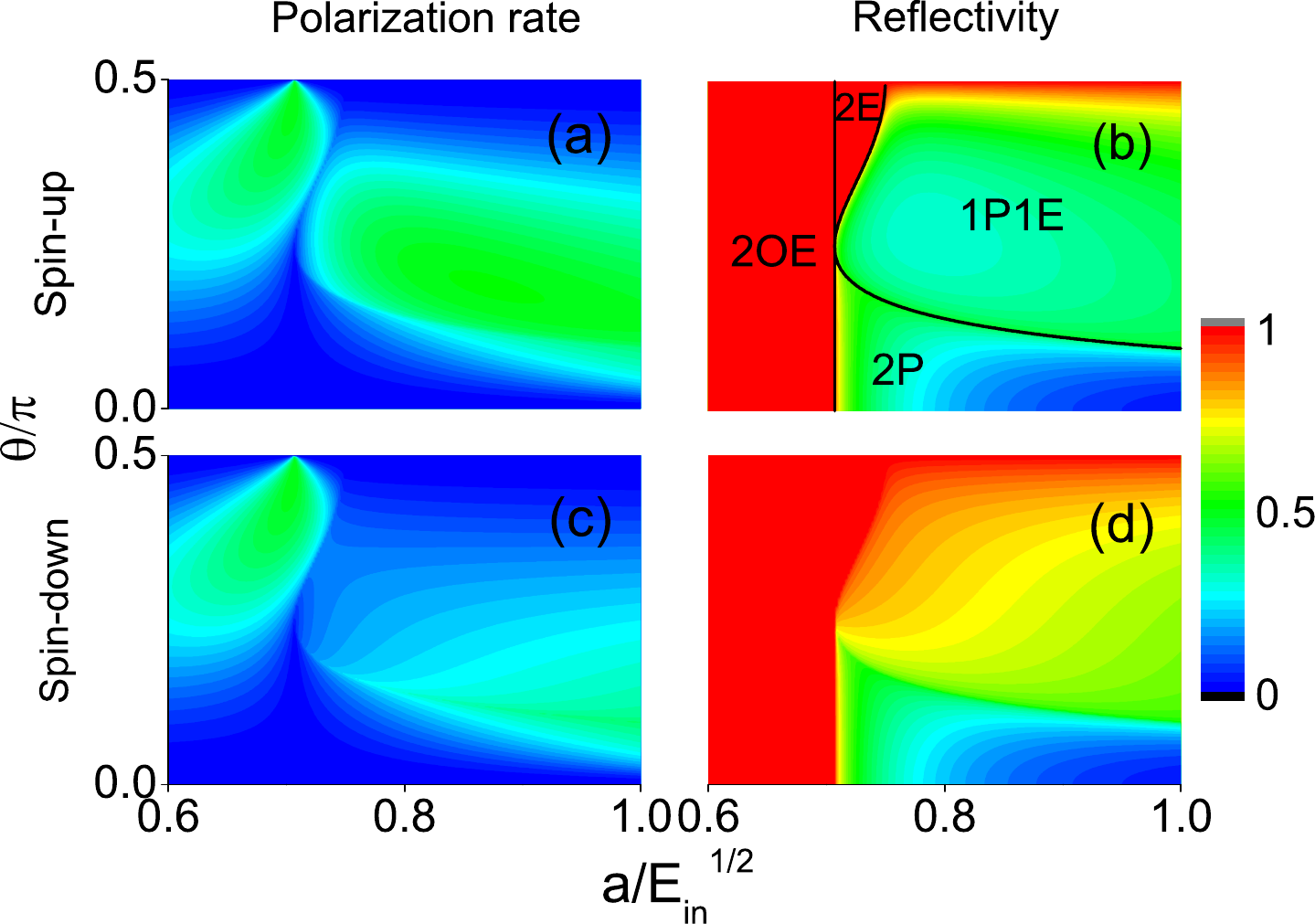}\caption{{\protect\footnotesize (Color
online) The contour plot of the spin-polarization efficiency and reflectivity
versus the incident angle and the SO-coupling strength }$a/\sqrt{E_{in}}%
${\protect\footnotesize . The potential barrier }$V_{0}=1.5E_{in}%
${\protect\footnotesize . (a) and (b) are for the incident atoms prepared in
the spin-up hyperfine state; while (c) and (d) are for the spin-down. The four
parameter regimes discussed in the text are labeled in (b), which are the same
for the other three subplots.}}%
\label{fig_polarization efficiency}%
\end{figure}

We calculate $r_{s}$ and $P_{s}$ as functions of incident angle and
SO-coupling strength $a$ with the atomic incident energy $E_{in}$ fixed. The
potential barrier height are set as $V_{0}=1.5E_{in}$. The results are shown
in Fig. \ref{fig_polarization efficiency}. Both the case of initially incident
spin-up and down atoms are considered. The four regimes referred above can be
clearly observed (also indicated in Fig. \ref{fig_polarization efficiency}(b)):

(i) When the SO-coupling strength is relatively small, i.e., $a<\sqrt
{V_{0}-E_{in}}$, inside the barrier the atomic beam can only exist in the form
of oscillating evanescent wave (2OE) and hence it experiences total internal
reflection. This means that the atom mirror will have the property of
omni-reflection, i.e., atoms with any incident angle will be reflected. As
shown in Fig. \ref{fig_polarization efficiency} (a) and (c), in this regime
various spin-polarization rate can be achieved via adjusting the incident
angle $\theta$, and one can reach high polarization rate (approximately up to
$0.5$) for large incident angle, which is ideal for a spin-selective atom
mirror setup.

(ii) In a small parameter region of $a>\sqrt{V_{0}-E_{in}}$, when the atomic
incident angle exceeds the critical value $\theta_{c} \equiv\sin^{-1} \left[
\left(  a+\sqrt{a^{2}+E_{in}-V_{0}}\right)  /\sqrt{E_{in}}\right]  $, total
internal reflection can still take place, in this case only evancescent wave
solutions exist inside the barrier (2E). Similar to the previous case, the
atom mirror can also operate in this regime with high efficiency and high
polarization rate can be achieved.

\begin{figure}[h]
\includegraphics[width=8cm]{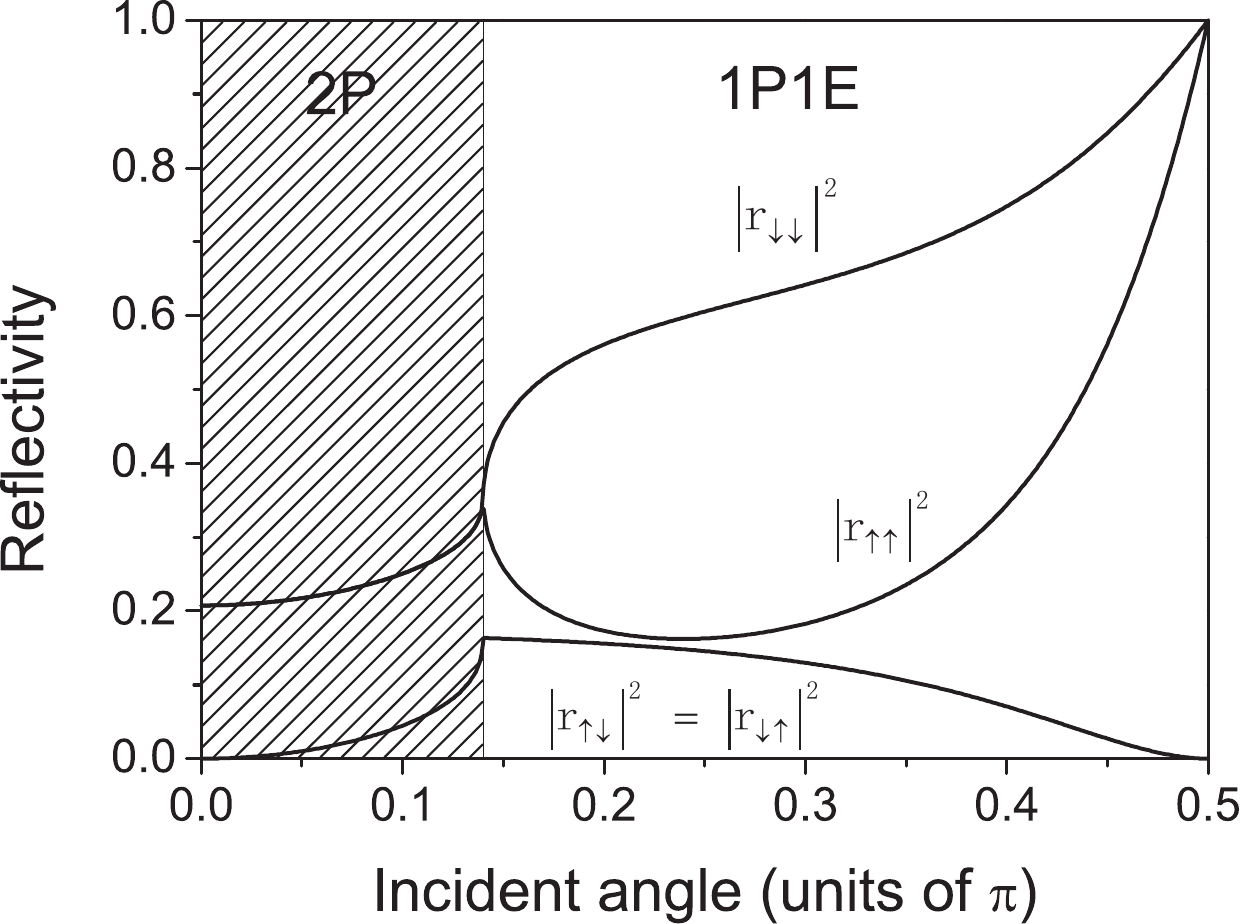}\caption{{\protect\footnotesize Reflectivity
versus incident angle. The parameters are set as }$a/\sqrt{E_{in}}=0.8$
{\protect\footnotesize and }$V_{0}=1.5E_{in}${\protect\footnotesize . Inside
the shadow area the atoms propagate inside the barrier in the form of two
propagating states (2P), while in the rest area one have one propagating state
and one evanescent state (1P1E) inside the barrier.}}%
\label{fig_1p1e}%
\end{figure}

(iii) With the increase of SO-coupling strength $a$, one enters into the 2P
regime when the incident angle $\theta$ is relatively small ($\theta<\sin
^{-1}\left[  \left(  a-\sqrt{a^{2}+E_{in}-V_{0}}\right)  /\sqrt{E_{in}%
}\right]  $). In this parameter region, most incident atoms will propagate
through the barrier in the form of propagating waves, which lead to very low reflectivity.

(iv) In other regions of $a>\sqrt{V_{0}-E_{in}}$, one enters into the 1P1E
regime in which one propagating wave solution and one evanescent wave solution
exist inside the barrier. In this regime part of the atoms propagate inside
the barrier and will not be reflected. The polarization rate will be
accordingly varied with the incident angle, and the maximum value of $0.5$ can
also be achieved.

By comparing the spin-polarization rate and reflectivity in Fig.
\ref{fig_polarization efficiency} for incident spin-up and spin-down atoms,
one can notice that they are identical except in the 1P1E regime. More
specifically, we have $\left\vert r_{\uparrow\downarrow}\right\vert
^{2}=\left\vert r_{\downarrow\uparrow}\right\vert ^{2}$ with $\left\vert
r_{\uparrow\uparrow}\right\vert ^{2}\neq\left\vert r_{\downarrow\downarrow
}\right\vert ^{2}$ in this regime. The results are shown in Fig.
\ref{fig_1p1e}. So in this parameter region the spin of reflected atoms can be
effectively polarized for an incident atom beam with equal spin-up and
spin-down populations.

We would like to note that under the present setup, for normal incidence
($\theta=0$), no spin polarization takes place. The spin-polarization
efficiency is also very low for small incident angle. This restricts the
working threshold of the atom mirror acting as a spin polarizer. However this
provides new possibilities for the realization of spin filter or
spin-selective atom mirror. For example, suppose that a mixture of
spin-$\uparrow$ and $\downarrow$ atoms incident upon the atom mirror at
relativity small incident angle (approaching normal incidence), and by
applying a homogeneous magnetic field along the $z$-direction one can induce a
Zeeman energy splitting between the atomic magnetic sub-levels. In this case
the essential physics inherent in the dispersion curve displayed in Fig.
\ref{fig_energy} would not be altered, however the incident energy of
spin-$\uparrow$ and $\downarrow$ atoms will be different. By appropriately
tuning the Zeeman energy splitting one can have, for example, spin-$\uparrow$
atoms in the 2OE regime while the spin-$\downarrow$ ones are in the 2P regime.
As a result only the spin-$\uparrow$ atoms will be reflected and the
spin-$\downarrow$ ones will penetrate through the potential and the
spin-selective reflection can then be realized. The quantum-state-selective
mirror reflection of atoms have been studied before in \cite{state-selective
atomic mirror}, in which the evanescent field formed at a dielectric interface
serves as the atom mirror. Here we provide a new scheme to construct this
special atom optical element.

\section{case of equal rashba-dresselhaus spin-orbit interaction}

\begin{figure}[h]
\includegraphics[width=8cm]{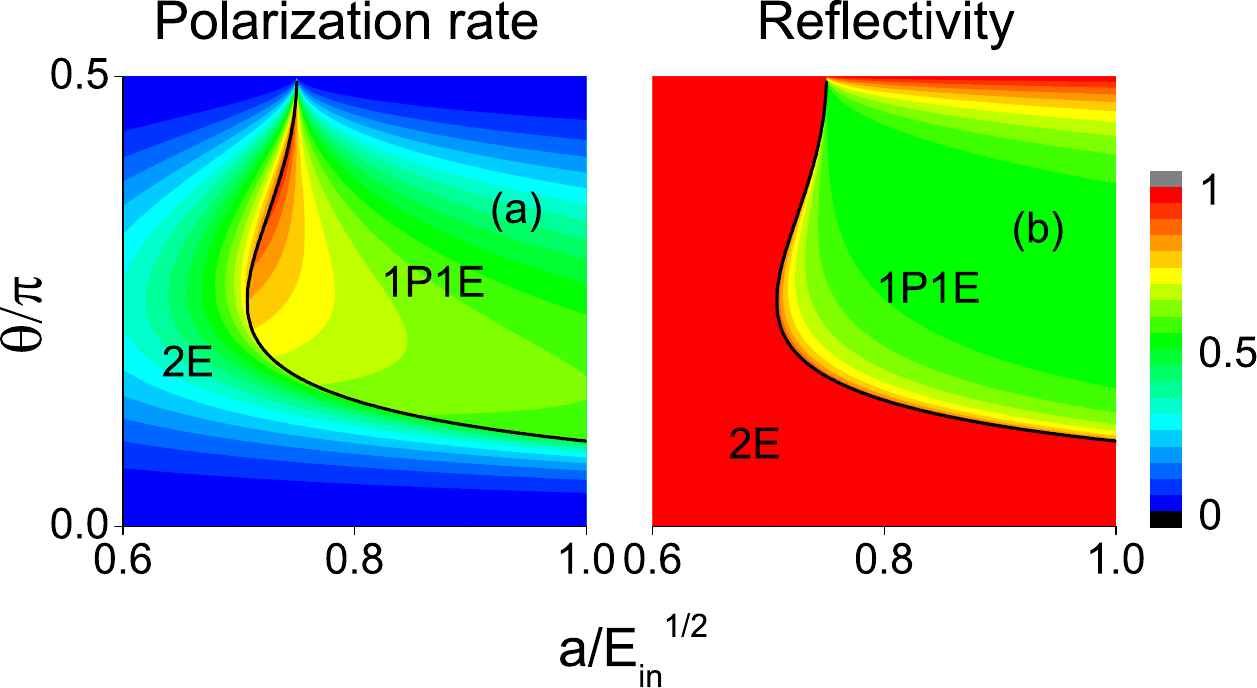}\caption{{\protect\footnotesize (Color
online) Same as Fig. \ref{fig_polarization efficiency} except that the
SO-coupling term have been replaced by }$k_{y}\sigma_{x}$%
{\protect\footnotesize . The results are identical for incident spin-}%
$\uparrow${\protect\footnotesize and }$\downarrow$%
{\protect\footnotesize atoms. The black line separates the regime of 2E and
1P1E.}}%
\label{fig_pysx}%
\end{figure}

The very first SO-interaction implemented in neutral cold atoms is of
one-dimensional equal Rashba-Dresselhaus type \cite{1D-SO}, so here we briefly
address the case of replacing the Rashba SO-interaction in our model by the
one-dimensional one. First consider the case that the Hamiltonian inside the
potential barrier is%
\begin{equation}
H=\frac{\hbar^{2}\boldsymbol{k}^{2}}{2m}+\frac{\hbar^{2} a}{m} k_{y}\sigma
_{x}+V_{0}. \label{eq_hamiltonian2}%
\end{equation}
The dispersion equation then becomes%
\begin{equation}
\left(  \boldsymbol{k}^{2}+V_{0}-E_{k}\right)  ^{2}-4a^{2}k_{y}^{2}=0.
\label{eq_dispersion2}%
\end{equation}
A direct result of the dispersion relation (\ref{eq_dispersion2}) is that it
does not support oscillating evanescent eigenstate solutions. In addition to
that, in contrast to Eq. (\ref{eq_boundary}) the SO-coupling $k_{y}\sigma_{x}$
have no contribution to the boundary condition in the $x$-direction. As a
result, the reflection properties of the system change qualitatively. The
results are shown in Fig. \ref{fig_pysx}.

For the parameters chosen here ($V_{0}>E_{in}$), one cannot enter
into the 2P regime. It is interesting to note that on the boundary which
separates the regime of 2E and 1P1E, maximum polarization rate ($P\approx1$)
can be achieved.

Finally, we comment on the case for which the SO-coupling term is propotional
to $k_{x}\sigma_{y}$, we found out that evanescent solutions will be absent
and atomic spin cannot be polarized upon reflection.

\section{summary}

In summary, we have studied the possibility of creating a spin-sensitive atom
mirror via SO-coupling. The spin polarization rate and reflectivity are
calculated as a function of incident energy, incident angle as well as the
SO-coupling strength. Depending on these parameters, incident atoms will
subject to quite different scattering process. We carefully analyzed these
results and showed that the atom mirror can effectively polarize the atomic
spin upon reflection. Due to the rich spin-dependent scattering properties
inherent in this system, the atom mirror can also perform spin-selective
reflection. These properties can find applications in the spin-dependent atom
interferometer and quantum measurement. In experiment, one can expect that the
predicted effect can be readily measured from the density evolution of the
atomic ensemble via absorption imaging \cite{detect}.

\begin{acknowledgments}
We thank Han Pu for careful reading and many useful comments on the
manuscript. This work is supported by National Key Research Program of China
under Grant No. 2016YFA0302000 and Grant No. 2011CB921604, National Natural
Science Foundation of China under Grant No. 11374003.
\end{acknowledgments}

\end{document}